\documentclass[aps,reprint,superscriptaddress,showpacs,prb]{revtex4-1} 

\usepackage{amsmath}
\usepackage{amssymb}
\usepackage{graphicx}
\usepackage{bm}       % provides bold math symbols via \bm{}
\usepackage{hyperref} % create hyperlinks within the document
\hypersetup{
     colorlinks=true,       % false: boxed links; true: colored links
     linkcolor=blue,        % color of internal links (change box color with linkbordercolor)
     citecolor=blue,        % color of links to bibliography
     filecolor=blue,        % color of file links
     urlcolor=cyan          % color of external links
}

\usepackage[usenames,dvipsnames]{xcolor}

\newcommand{\Egap}{\ensuremath{E_\mathrm{gap}}}
\newcommand{\Eimp}{\ensuremath{E_\mathrm{imp}}}
\newcommand{\muc}{\ensuremath{\mu_\mathrm{c}}}
\newcommand{\QDIII}{\ensuremath{{\cal Q}_\mathrm{DIII}}}
\newcommand{\QTRIM}{\ensuremath{{\cal W}}}

\newcommand{\QZZ}{\ensuremath{{\cal Q}}}
\newcommand{\Tc}{\ensuremath{T_\mathrm{c}}}
\newcommand{\U}{\ensuremath{u}}
\newcommand{\Uc}{\ensuremath{\U_\mathrm{c}}}
\newcommand{\Ztwo}{\ensuremath{\mathbb{Z}_2}}

\newcommand{\braket}[3]
{\ensuremath{\left\langle {#1} \left| {#2} \right| {#3}
    \right\rangle}}
\newcommand{\ket}[1]{\ensuremath{| #1 \rangle}}

\renewcommand{\vec}[1]{\bm{ #1 }}
\DeclareMathOperator{\pf}{Pf}
\DeclareMathOperator{\sgn}{sgn}

\begin{document}
\title{Symmetry-protected topological invariant and Majorana impurity
  states in time-reversal-invariant superconductors}

\author{Lukas Kimme} \affiliation{Institut f\"ur Theoretische
Physik, Universit\"at Leipzig, D-04103 Leipzig, Germany}
\author{Timo Hyart} \affiliation{Instituut-Lorentz, Universiteit
Leiden, Post Office Box 9506, 2300 RA Leiden, The Netherlands}
\author{Bernd Rosenow} \affiliation{Institut f\"ur Theoretische
Physik, Universit\"at Leipzig, D-04103 Leipzig, Germany}

\date{June 2nd, 2015}

\begin{abstract}
  We address the question of whether individual nonmagnetic impurities
  can induce zero-energy states in time-reversal-invariant topological
  superconductors, and define a class of symmetries which guarantee
  the existence of such states for a specific value of the impurity
  strength. These symmetries allow the definition of a position-space
  topological \Ztwo\ invariant, which is related to the standard bulk
  topological \Ztwo\ invariant. Our general results are applied to the
  time-reversal-invariant $p$-wave phase of the doped
  Kitaev-Heisenberg model, where we demonstrate how a lattice of
  impurities can drive a topologically trivial system into the
  nontrivial phase.
\end{abstract}

\maketitle

Local impurities in superconductors (SCs) give rise to astonishing
physics.  \cite{Abrikosov1960, Balatsky2006, Alloul2009, Hudson2001,
  Sau2012, Hu2013, Maki1999, Khaliullin1997} Magnetic impurities in
$s$-wave SCs lead to pair breaking, and can induce a quantum phase
transition to a metallic state with gapless superconductivity near the
transition point.\cite{Abrikosov1960} Due to Anderson's theorem,
nonmagnetic impurities have little influence on $s$-wave
SCs.\cite{Anderson1959} However, in unconventional SCs, where the sign
of the order parameter depends on the direction of momentum,
scattering by impurities leads to pair-breaking since the momentum
direction of the paired electrons is changed without changing the
phase.  \cite{Balatsky2006,Alloul2009} Thus, impurities give rise to
subgap states and can be used to probe high-\Tc\
superconductivity.\cite{Balatsky2006, Alloul2009, Hudson2001}

Here, we focus on impurity bound states in time-reversal (TR)
invariant odd-parity SCs.  These SCs belong to symmetry class DIII of
the Altland-Zirnbauer classification \cite{Altland1997} and come in
two variants, characterized by a \Ztwo\ topological invariant \QZZ.
\cite{Schnyder2008, Kitaev2009, Qi2009, Hasan2010, Qi2011, Sato2009,
  Fu2010, footnote:invariant3D} The topologically nontrivial SC has
protected Majorana boundary modes.  It turns out that \QZZ\ also
predicts the pattern of ground-state degeneracies on a torus, when
switching between periodic and antiperiodic boundary conditions.
\cite{Read2000} Denoting a pair of states $(\ket{\psi}, T \ket{\psi})$
related by time reversal $T$ as a Kramers pair, ground states are
different depending on whether the number of unpaired Kramers pairs
below the Fermi level is even or odd, designated in the following as
even or odd ``Kramers parity.''  Single-band odd-parity SCs have
$\Delta(-\vec{k}) = - \Delta (\vec{k})$,\cite{Sato2009} hence their
order parameter vanishes at all TR-invariant momenta (TRIM) $\vec{K}$
with $\vec{K} = -\vec{K}$ up to reciprocal lattice vectors, such that
for each TRIM below the Fermi level there is one unpaired Kramers
pair.  The Kramers parity is thus determined by the number of TRIM
enclosed by the Fermi surface, and odd-parity SCs where this number is
odd are topologically nontrivial.  \cite{Sato2009,Fu2010}

Zero-energy bound states in SCs are intriguing Majorana states.
\cite{Read2000,Kitaev2001,Ivanov2001,Wilczek2009} Thus, it may be
interesting to artificially create them by tuning an impurity
potential, but it is also important to understand how to avoid
accidental zero-energy states from nonmagnetic disorder, which may
interfere with protocols using protected Majorana zero-energy
states,\cite{Fulga2013} occurring for instance in the center of a
vortex.\cite{Kopnin1991, Volovik1999} In this paper, we derive
conditions for the existence of zero-energy impurity states in
TR-invariant SCs.  To this end, we deduce conditions for the existence
of a position-space topological invariant \QDIII, which for gapped
translationally invariant systems is equivalent to \QZZ\ and the
Kramers parity.  We show that upon introduction of a local impurity
potential into the system, the conditions for the existence of \QDIII\
also guarantee the emergence of zero-energy impurity bound states for
a suitably chosen impurity strength.  In particular, we find that the
existence of symmetries protects zero-energy impurity bound states,
such that disorder may introduce states with energies less than the
thermal energy even at low temperatures.  When an impurity bound state
moves through the Fermi level, it changes the Kramers parity and
\QDIII\ but not \QZZ, since it is spatially localized and insensitive
to boundary conditions.  However, a lattice of impurities hosts
extended states, and we show that partially moving such an impurity
band through zero energy can, for a broad range of potential
strengths, turn a topologically trivial SC into a nontrivial one.

\emph{Model:}~We consider a general TR-invariant Bogoliubov--de
Gennes Hamiltonian in symmetry class DIII \cite{Altland1997} for an
$N$-site lattice in the position-space basis
\begin{equation}
  H = \frac{1}{2} \left( c^\dagger, c\right) {\cal H} \begin{pmatrix}
    c\\ c^\dagger \end{pmatrix},\quad 
  {\cal H} = \begin{pmatrix}
    h & \Delta\\
    -\Delta^\star & -h^T \end{pmatrix} , 
    \label{eq:BdGHamiltonian}
\end{equation}
where $c=(c_\uparrow, c_\downarrow)$,
$c_\sigma = \left(c_{1, \sigma},\dots, c_{N, \sigma}\right)$, and
$c_{i,\sigma}$ annihilates a fermion with spin $\sigma$ on site $i$.
Hermiticity of the Hamiltonian and Fermi statistics requires
$h=h^\dagger$, $\Delta=-\Delta^T$.  Hamiltonians in DIII obey both the
particle-hole (PH) symmetry $\{P,{\cal H}\}=0$, $P = \tau_1K$ and TR
symmetry $[T,{\cal H}]=0$, $T=i\sigma_2K$.  Here $\vec{\tau}$ and
$\vec{\sigma}$ denote the Pauli matrices in PH and spin space,
respectively, and $K$ is the operator of complex conjugation.
Together, these symmetries give rise to the chiral symmetry
$\{C,{\cal H}\}=0$,
$C = i P T = \tau_1\otimes\sigma_2$.\cite{Schnyder2008} Hence, every
eigenvector $\ket{\psi}$ with energy $E$ has a Kramers partner
$T\ket{\psi}$ with energy $E$, a PH partner $P\ket{\psi}$, and a
``chiral'' partner $C\ket{\psi}$ both with energy $-E$.

\begin{figure}
  \includegraphics[width=\columnwidth]{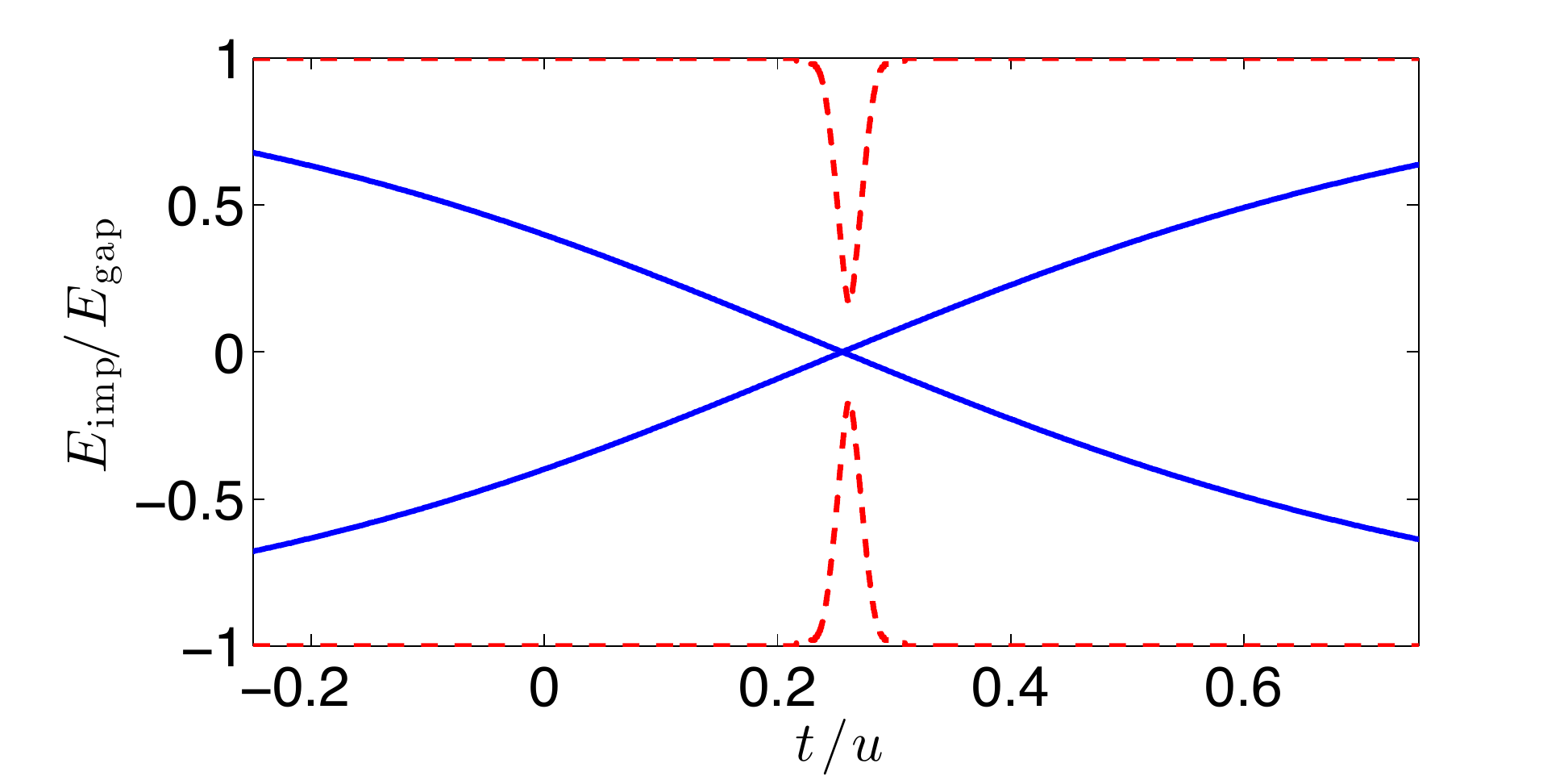}
  \caption{\label{fig:subgapstate} Two prototypical behaviors of the
    energy of an impurity state $\Eimp/\Egap$ as a function of the
    inverse impurity strength $t/\U$.  The solid blue line shows a
    symmetry protected zero-energy crossing, whereas the dashed red
    line shows an avoided crossing, because the symmetry is absent.
    The relevant symmetries are listed in
    Table~\ref{tab:SymmetriesUnified}. Both systems are in the
    topologically non-trivial phase.  The blue curve is computed for
    the TR invariant $p$-wave phase of the doped Kitaev-Heisenberg
    model (parameters: $\mu=1.3\,t,\eta=0.05\,t$); for the red curve
    anisotropic Rashba spin-orbit coupling with
    $(\kappa_x, \kappa_y, \kappa_z) = (0,1,2)$ and
    $\lambda_\mathrm{R}=0.89\,\eta$ was added.}
\end{figure}

We describe a local nonmagnetic impurity at site $i_0$ by the
Hamiltonian
\begin{equation}
  H(\U) = H+H_\mathrm{imp}(\U),\quad
  H_\mathrm{imp}(\U) = \U \sum_\sigma c_{i_0,\sigma}^\dagger
  c_{i_0,\sigma} .
  \label{eq:onsitepotential}
\end{equation}
\emph{Results:}~To get insight into the existence of zero-energy
impurity states, we note that in the absence of superconductivity
$H_0(u) = H(u, \Delta=\bm{0}_{2N})$ has a zero-energy eigenvalue for a
critical impurity strength $\Uc^0$.\cite{footnote:classAII} Without
accidental degeneracies, the zero-energy eigenspace is spanned by the
mutually orthogonal states $\ket{\psi^0},T\ket{\psi^0},P\ket{\psi^0}$
and $C\ket{\psi^0}$.  We now ask whether these states are split by a
superconducting coupling $H_\Delta = H - H_0$ in
first-order-degenerate perturbation theory, and argue that such a
splitting is evidence for an avoided crossing, and thus the absence of
a zero-energy state in the full problem.  Due to TR and PH symmetry,
$H_\Delta$ cannot couple $\ket{\psi^0}$ to $T\ket{\psi^0}$ or
$P\ket{\psi^0}$, but the coupling to $C\ket{\psi^0}$ is finite in
general and leads to an energy splitting.  \cite{SupplementalMaterial}
However, in the presence of a unitary symmetry $U$, which commutes
with $H_0(u)$ and $H_\Delta$ and anticommutes with $C$, the coupling
between $\ket{\psi^0}$ and $C\ket{\psi^0}$ vanishes: since
$U\ket{\psi^0} = \lambda\ket{\psi^0}$ with $|\lambda|=1$, we find that
$\braket{\psi^0}{H_\Delta C}{\psi^0} = \braket{\psi^0}{U^\dagger
  H_\Delta CU}{\psi^0}$,
and from $\{H_\Delta C,U\}= 0$ it follows that
$\braket{\psi^0}{H_\Delta C}{\psi^0} = -\braket{\psi^0}{H_\Delta
  C}{\psi^0}$.
Consequently, $\braket{\psi^0}{H_\Delta C}{\psi^0}$ vanishes, and
there is no energy splitting.  This fundamental impact of such a
symmetry $U$ on the energy $\Eimp$ of the impurity bound state is
illustrated in Fig.~\ref{fig:subgapstate}.  There we depict
$\Eimp(u^{-1})$ obtained from $T$-matrix\cite{Balatsky2006}
calculations for two models: first for the doped Kitaev-Heisenberg
(KH) model,\cite{Kitaev2006, Giniyat0910} which, as we will
demonstrate, has additional symmetries protecting the zero-energy
crossings, and second for the case where we added to this model Rashba
spin-orbit coupling and modified the order parameter $\Delta(\vec{k})$
in order to break all these symmetries.

In order to understand the existence of zero-energy states in the full
problem, we note that the determinant $\det[{\cal H}(\U)]$ can be
expressed as a product of the eigenvalues of ${\cal H}(\U)$. Thus, if
the system without impurity is gapped, a zero of $\det[{\cal H}(\U)]$
for a critical impurity strength $\Uc$ indicates the existence of a
zero-energy impurity bound state.  As ${\cal H}(\U)$ is local in \U,
and since there is a spin and particle-hole degree of freedom at each
lattice site, one finds that $\det[{\cal H}(\U)]$ is a fourth-order
polynomial in \U.  For a general Hamiltonian in class DIII, it is
difficult to determine under which conditions this polynomial has
zeros for a real-valued impurity strength $\Uc$.  In the following, we
reduce the problem to the analysis of a first-order polynomial by
considering the Pfaffian of redundant subblocks of ${\cal H}$.  This
will allow us to show non-perturbatively that the presence of a
symmetry with $[{\cal H},U]=0$ and $\{C,U\}=0$ indeed ensures the
existence of a zero-energy impurity bound state.

We first use the transformation $V = [\bm{1}_{4N} +
(i\tau_2)\otimes\sigma_2\otimes \bm{1}_N]/\sqrt{2}$, which
diagonalizes $C$, to bring ${\cal H}$ into a block off-diagonal form
\begin{equation}
  V^\dagger {\cal H} V = \begin{pmatrix}
    \bm{0}_{2N} & D \\ D^\dagger & \bm{0}_{2N}
  \end{pmatrix}\label{eq:HamiltonianBlockOffDiagonal} , 
\end{equation}
with $D \equiv h\sigma_2+\Delta = -D^T$.  Because $D$ is
antisymmetric, $\pf(D)$ exists and $ |\pf(D)|^4= \det{\cal H} $, such
that zero-energy eigenvalues of ${\cal H}$ occur whenever $\pf(D)=0$.
Since \U\ appears only in one entry in the upper and lower triangle of
the matrix $D(\U)$, respectively, $\pf[D(u)] = z(u-u_\mathrm{c})$ is a
linear complex function with $z,\Uc\in\mathbb{C}$.  If \Uc\ is real,
the complex phase of $\pf[D(u)]$ does not depend on \U\ and the system
is bound to have a single zero-energy crossing of Kramers pairs at
\Uc.  We stress that in general there is no reason for $\Uc$ to be
real, such that no value of the real control parameter \U\ would yield
zero-energy states. In the following, we will show that $\Uc$ is
indeed real provided that a symmetry of the Hamiltonian exists which
anticommutes with the chiral operator $C$.

Every possible unitary transformation $U$ satisfying $\{U,C\}=0$ has
the property \cite{SupplementalMaterial}
\begin{equation}
  V^\dagger U V = \begin{pmatrix}
    \bm{0}_{2N} & W \\
    W^\star & \bm{0}_{2N}
  \end{pmatrix}, \label{eq:generalSymmetryOperator}
\end{equation}
with $W$ unitary due to the unitarity of $U$ and $V$.  Provided that
$U$ is a symmetry of ${\cal H}$ with $[{\cal H},U]=0$ it follows that
\begin{equation}
  [\pf(D)]^\star = 
  \frac{(-1)^N}{\det W} \pf(D) . 
  \label{eq:PfaffianConstantPhase}
\end{equation}
Here, we first used the general properties
$[\pf(B)]^* = (-1)^N \pf(B^\dagger)$ and $\det(A)\pf(B)=\pf(ABA^T)$ of
the Pfaffian to write
$[\pf(D)]^\star = \frac{(-1)^N}{\det W} \pf(WD^\dagger W^T)$.  By
utilizing $WD^\dagger=DW^\star$, which is equivalent to the symmetry
condition $[{\cal H},U]=0$, and the unitarity of $W$, we then arrive
at Eq.~\eqref{eq:PfaffianConstantPhase}.  This equation implies that
$\sqrt{(-1)^N/ \det W} \pf(D(u)) $ is a real-valued function, and
therefore $\Uc$ is real.  This demonstrates that in the presence of a
symmetry $U$ the existence of the zero-energy states is guaranteed for
a suitably chosen impurity strength $\Uc$.

To get some intuition about possible symmetries, we first specialize
to a situation where $U$ can be decomposed into a product
$U=\tau_\mu\otimes\sigma_\nu\otimes R$ of an internal transformation
$\tau_\mu\otimes\sigma_\nu$ and a lattice transformation $R$, which
satisfies $R^T = R^{-1}$ as it is a permutation of lattice
sites. Then, the condition $\{U,C\}=0$ implies that not all 16
combinations $\tau_\mu \otimes \sigma_\nu$ can be used to construct
symmetries $U$, but only the eight combinations listed in
Table~\ref{tab:SymmetriesUnified}.  Next, we expand
$h=\sum_{\nu=0}^3\sigma_\nu\otimes h_\nu$ into a spin-independent
single-particle part $h_0$ and spin-orbit couplings $h_1$, $h_2$,
$h_3$, and decompose $\Delta=i\sum_{\nu=0}^3 \sigma_\nu
\sigma_2\otimes d_\nu$ into a singlet component $d_0$ and triplet
components $d_1$, $d_2$, $d_3$. Then, for every allowed choice of
$\tau_\mu \otimes \sigma_\nu$, a subset of the $h_\nu$, $d_\nu$
anticommutes with $R$, and the remaining $h_\nu$, $d_\nu$ commute with
$R$; see Table~\ref{tab:SymmetriesUnified}. In the particularly simple
case where $U$ does not contain a lattice transformation,
i.e., $R\equiv \bm{1}_N$, the anticommutation condition $\{\, \cdot \,
,R\}=0$ implies that the respective $h_\nu$, $d_\nu$ vanish
identically, whereas the commutation relation $[\, \cdot \, ,R]=0$ is
trivially satisfied.

Now we are in a position to treat the special case of impurity bound
states in spin-polarized SCs (belonging to symmetry class D
\cite{Altland1997}) as a first application of our formalism. The
specific choice $U= \tau_0 \otimes \sigma_3 \otimes \bm{1}_N$ implies
that the matrices $h_1$, $h_2$, $d_0$, $d_3$, which couple up and down
spins, have to vanish; see first row in
Table~\ref{tab:SymmetriesUnified}.  Then, the Hamiltonian matrix
decomposes into two uncoupled blocks
${\cal H} = {\cal H}^\uparrow \oplus {\cal H}^\downarrow$, related by
TR symmetry ${\cal H}^\downarrow = T {\cal H}^\uparrow T^{-1}$.  Each
of the blocks ${\cal H}^\sigma$ is not TR symmetric but still obeys PH
symmetry and thus can be an arbitrary member of symmetry class D.
From our analysis it follows that ${\cal H}^\uparrow$ hosts a
zero-energy impurity bound state for a suitably chosen impurity
strength while ${\cal H}^\downarrow$ provides its Kramers partner.
This generalizes the result for $p$-wave SCs obtained in
Ref.~\onlinecite{Sau2012} to arbitrary spin-polarized SCs in all
spatial dimensions. The symmetries in rows two and three of
Table~\ref{tab:SymmetriesUnified} imply a decomposition into two class
D blocks as well, with spins polarized in the $y$ and $x$ directions,
respectively.

The symmetry $U=\tau_3\otimes\sigma_0\otimes\bm{1}_N$ in the fourth
row of Table~\ref{tab:SymmetriesUnified} requires the absence of
superconductivity.  Hence, the coupling between the particle and the
hole-sector vanishes, and the Hamiltonian decomposes into two spin-1/2
TR-invariant systems belonging to symmetry class AII.
\cite{Altland1997} Thus, we have shown that every gapped system in AII
hosts zero-energy impurity bound states for a suitably chosen impurity
strength. The last four rows of Table~\ref{tab:SymmetriesUnified} are
formally obtained by multiplying the first four rows with the chiral
operator $C$. In the context of electronic SCs, there is no obvious
example for their use.

More generally, $R\neq \bm{1}_N$, and the symmetry $U$ realizes a
combination of a lattice transformation and a rotation in spin and
particle-hole space which is required to keep a spin-orbit coupling
$L\cdot S$ of angular momentum and spin invariant. An important
example are spatial reflections about a mirror plane, accompanied by
the appropriate spin rotation.\cite{Ueno2013, Zhang2013, Chiu2013,
  Marimoto2013} We discuss specific examples for such symmetries in
the context of the doped KH model.

The presence of a symmetry $U$ is sufficient but not necessary for the
existence of zero-energy impurity states.  There are conditions not
related to symmetries for which $\pf(D)$ has a real zero for some
impurity potential.\cite{SupplementalMaterial} However, while such
conditions can be satisfied in single-particle Hamiltonians, they are
expected to be less robust than symmetry conditions when the
single-particle Hamiltonian is obtained from a self-consistent mean
field approximation to an interacting Hamiltonian which already
includes the impurity potential.

\begin{table}
  \caption{ We list all eight types of unitary
    symmetry operators of the form
    $U=\tau_\mu\otimes\sigma_\nu\otimes R$ with $R^T=R^{-1}$ which satisfy
    $\{U,C\}=0$ and hence guarantee the existence of zero-energy
    impurity bound states.  The symmetry condition $[{\cal
      H},U]=0$ implies that the matrices $h_\nu,\,d_\nu$ defined by the expansions
    $h=\sum_{\nu=0}^3\sigma_\nu\otimes h_\nu$,
    $\Delta=i\sum_{\nu=0}^3 \sigma_\nu \sigma_2\otimes d_\nu$ are
    restricted by (anti)commutation relations with $R$.  Namely the
    $h_\nu,d_\nu$ listed in the second [third] column have to anticommute
    [commute] with $R$.  $R=\bm{1}_N$ implies that matrices in the
    second [third] column vanish [are unrestricted]. }
  \label{tab:SymmetriesUnified}
  \begin{ruledtabular}
  \begin{tabular}{ccc}
    $U$ & $\{\, \cdot \, ,R\}=0$ & $[\, \cdot \, ,R]=0$ \\\hline
    $\tau_0\otimes\sigma_3\otimes R$ & $h_1,h_2,d_0,d_3$ & $h_0,h_3,d_1,d_2$\\
    $\tau_3\otimes\sigma_2\otimes R$ & $h_1,h_3,d_0,d_2$ & $h_0,h_2,d_1,d_3$\\
    $\tau_0\otimes\sigma_1\otimes R$ & $h_2,h_3,d_0,d_1$ & $h_0,h_1,d_2,d_3$\\
    $\tau_3\otimes\sigma_0\otimes R$ & $d_0,d_1,d_2,d_3$ & $h_0,h_1,h_2,h_3$\\
    \hline
    $C(\tau_0\otimes\sigma_3\otimes R)$ & $h_0,h_3,d_1,d_2$ & $h_1,h_2,d_0,d_3$\\
    $C(\tau_3\otimes\sigma_2\otimes R)$ & $h_0,h_2,d_1,d_3$ & $h_1,h_3,d_0,d_2$\\
    $C(\tau_0\otimes\sigma_1\otimes R)$ & $h_0,h_1,d_2,d_3$ & $h_2,h_3,d_0,d_1$\\
    $C(\tau_3\otimes\sigma_0\otimes R)$ & $h_0,h_1,h_2,h_3$ & $d_0,d_1,d_2,d_3$\\
  \end{tabular}
  \end{ruledtabular}
\end{table}

Exploiting the constant phase of $\pf(D)$ in the presence of a
symmetry $U$, we define a topological invariant
$\QDIII = \sgn[\sqrt{(-1)^N/ \det W} \pf(D)]$, which changes whenever
one Kramers pair crosses the Fermi energy.  To establish a connection
between \QDIII\ and the widely used bulk topological invariant \QZZ\
for translationally invariant odd-parity single-band SCs, we define
$D(\vec{k}) = h(\vec{k})\sigma_2 + \Delta(\vec{k})$ for each momentum
$\vec{k}$ in analogy to Eq.~\eqref{eq:HamiltonianBlockOffDiagonal}.
For a TRIM $\vec{K}$, $\Delta(\vec{K})=\bm{0}_2$ and
$h(\vec{K})=\xi(\vec{K})\sigma_0$, where $\xi(\vec{K})$ is the
single-particle energy with respect to the Fermi energy.  Hence,
$D(\vec{K})$ is antisymmetric and in agreement with Sato:
\cite{Sato2009}
\begin{gather}
  \QZZ = \prod_{\vec{K}\in\mathrm{TRIM}}
  \QTRIM(\vec{K}), \label{eq:DefinitionQZZ}
\end{gather}
where
$\QTRIM(\vec{K})\equiv\sgn[i\pf D(\vec{K})] = \sgn \xi(\vec{K})$, so
that \QZZ\ counts the number parity of TRIM below the Fermi level and
thus the Kramers parity.  Consequently, $\QDIII=\QZZ$ for these
systems.\cite{footnote:gaplessTopSC} It is straightforward to
generalize our definitions to multiband SCs as well. We will make use
of this generalization to demonstrate that a lattice of impurity
states can drive a SC into a topologically nontrivial phase.

\begin{figure}[t]
  \includegraphics[width=\columnwidth]{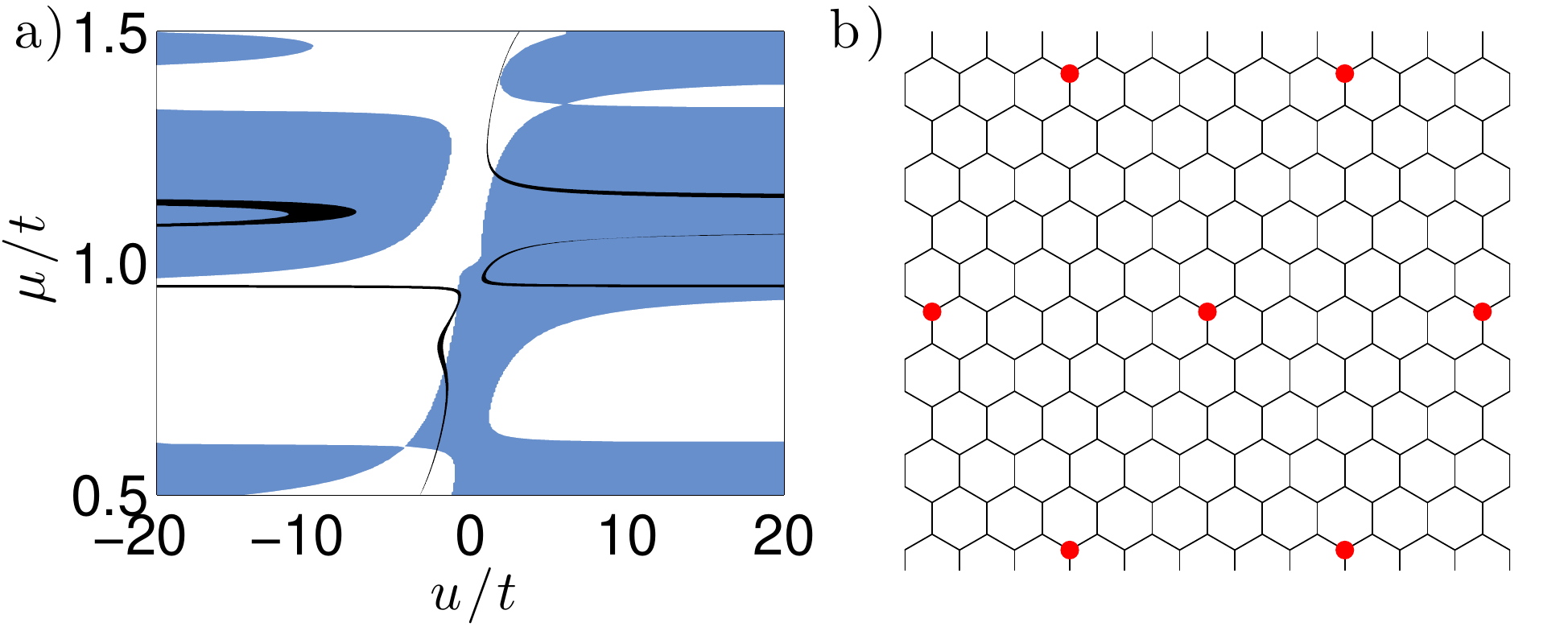}
  \caption{\label{fig:phaseDiagram}(a) Phase diagram of \QZZ\ for an
    impurity lattice with impurity distance $a_\mathrm{imp}=5$ in the
    TR-invariant $p$-wave phase of the doped KH model as a function of
    the impurity strength \U\ and the chemical potential $\mu$.  Blue
    denotes the topologically trivial phase $\QZZ=+1$ whereas white
    denotes the nontrivial phase $\QZZ=-1$.  Black denotes regions
    where the system is gapless.\protect\cite{footnote:gaplessTopSC}
    (b) Impurity lattice for $a_\mathrm{imp}=5$; red dots mark impurity
    sites.}
\end{figure}

\emph{Impurities in the doped KH model:} We illustrate our general
results by applying them to the TR-invariant $p_x \pm ip_y$-wave phase
of the doped KH model on the honeycomb lattice, \cite{Kitaev2006,
  Giniyat0910,Hyart2012,Okamoto2012,Scherer2014} which is paradigmatic
for a number of interesting topological phases.\cite{Hyart2014} This
phase is a two-dimensional analog of the $B$ phase of superfluid
$^3$He and undergoes a topological phase transition at a critical
value of the chemical potential.\cite{Hyart2012} Consider, therefore,
the mean-field Hamiltonian
\begin{eqnarray}
  H_\mathrm{KH} &=& -\mu \sum_{\vec{k},s,\sigma}
  f_{\vec{k},s,\sigma}^\dagger f_{\vec{k},s,\sigma} -
  \sum_{\vec{k},\sigma} [t(\vec{k})f_{\vec{k},1,\sigma}^\dagger
  f_{\vec{k},2,\sigma} +\mathrm{h.c.}]  \nonumber \\ && \hspace{-0.4cm}+
  \sum_{\vec{k},\sigma} \left\{[-\sigma d^x(\vec{k})+id^y(\vec{k})]
    f_{\vec{k},1,\sigma}^\dagger f_{-\vec{k},2,\sigma}^\dagger +
    \mathrm{h.c.}\right\}
  \label{eq:KHHamiltonian}
\end{eqnarray}
where $f_{\vec{k},s,\sigma}$ annihilates a fermion with spin $\sigma$
on sublattice $s$, $\mu$ is the chemical potential,
$t(\vec{k}) = t(e^{i\vec{\delta}_x\cdot
  \vec{k}}+e^{i\vec{\delta}_y\cdot \vec{k}}+e^{i\vec{\delta}_z\cdot
  \vec{k}})$
is the nearest-neighbor hopping, and
$d^x=3i\eta( e^{i\vec{\delta}_x\cdot \vec{k}} -
e^{i\vec{\delta}_y\cdot \vec{k}} )/\sqrt{6},
d^y=i\eta(e^{i\vec{\delta}_x\cdot \vec{k}}+e^{i\vec{\delta}_y\cdot
  \vec{k}} - 2 e^{i\vec{\delta}_z\cdot \vec{k}}) /\sqrt{2}, d^z=0$
are the components of the $\vec{d}$ vector describing $p_x\pm ip_y$
spin-triplet pairing; for small $\vec{k}$, $\vec{d}\sim (k_x,k_y,0)$.
Here, $\eta$ characterizes the superconducting gap and
$\vec{\delta}_{x,y,z}$ are the nearest-neighbor vectors.

In Eq.~\eqref{eq:KHHamiltonian} we chose the spin quantization axis
such that only equal-spin particles are paired; hence
$[{\cal H}_\mathrm{KH},\sigma_3]=0$, which is a nonspatial symmetry
protecting zero-energy states; cf. Table~\ref{tab:SymmetriesUnified}.
From the interacting Hamiltonian\cite{Giniyat0910} the $p$-wave phase
inherits symmetries acting on spin and spatial degrees of freedom.
\cite{You2012} Of these symmetries only the three mirror symmetries
$M_\gamma$ with respect to the $x$, $y$, or $z$ links satisfy
Eq.~\eqref{eq:generalSymmetryOperator}, for example
$M_z=\tau_0\otimes\sigma_1\otimes R_z$, where $R_z$ is the matrix for
the mirror permutation of the lattice sites with respect to a $z$
link.  Hence, also the $M_\gamma$ protect the zero-energy crossings
shown in Fig.~\ref{fig:subgapstate}.  It is instructive to add Rashba
spin-orbit coupling
$H_\mathrm{R} = i\lambda_\mathrm{R} \sum_{\langle ij
  \rangle,\alpha\beta} f_{i\alpha}^\dagger \left[ \kappa_\gamma \left(
    \vec{\sigma} \times \hat{\vec{\delta}}_\gamma \right)_z
\right]_{\alpha\beta} f_{j\beta}$,
with
$\hat{\vec{\delta}}_\gamma =
\vec{\delta}_\gamma/|\vec{\delta}_\gamma|$,
to the Hamiltonian while disregarding the effects that this coupling
would have if the order parameter was calculated self-consistently.
For $\lambda_\mathrm{R} \neq 0$ this breaks the non-spatial symmetry
$[{\cal H}_\mathrm{KH}+{\cal H}_\mathrm{R},\sigma_3]\neq 0$, but keeps
all spatial symmetries intact if $\kappa_\gamma=1$, $\gamma=x,y,z$.
Anisotropic Rashba coupling with $\kappa_z \neq 1$ breaks all mirror
symmetries except for $M_z$. By choosing different values for all
three $\kappa_\gamma$ one breaks all relevant symmetries and thus
avoids the impurity-induced zero-energy crossing.  This is illustrated
in Fig.~\ref{fig:subgapstate}.

In order to demonstrate that {\em extended} impurity states not only
change $\QDIII$ but also \QZZ, we consider a triangular lattice of
impurities with lattice constant $a_\mathrm{imp}=5$, amounting to an
impurity density of $2\%$ [see Fig.~\ref{fig:phaseDiagram}\,(b)]. We
calculate \QZZ\ by evaluating $\QTRIM(\vec{K})$ at the four TRIM as
well as the Chern number $C_\mathrm{imp}$ of each spin-resolved
impurity band formed by overlapping impurity subgap states, and
confirm that $\QZZ(u)=(-1)^{C_\mathrm{imp}}\QZZ(0)$.  Due to threefold
rotational symmetry of $H_\mathrm{KH}$\cite{You2012}
$\QTRIM(M)\equiv\QTRIM(M_1)=\QTRIM(M_2)=\QTRIM(M_3)\neq\QTRIM(\Gamma)$,
where $M_i$ denotes the $M$ points and $\Gamma$ denotes the origin of
the Brillouin zone.  $\QTRIM(M)$ as well as $\QTRIM(\Gamma)$ are the
sign of linear functions in $u$, respectively, and thus change
independently of each other at critical values $\Uc^M$ and
$\Uc^\Gamma$, respectively.  Hence, one can change
$\QZZ = \QTRIM(\Gamma)\QTRIM(M)$ by tuning \U.  In
Fig.~\ref{fig:phaseDiagram}\,(a) we show the phase diagram of \QZZ\
versus impurity strength \U\ and chemical potential $\mu$.  The clean
system is in the topologically trivial [nontrivial] phase for
$\mu<\muc\simeq 0.993\,t$ [$\mu>\muc$].  At each value of $\mu$ two
transitions occur at $\Uc^M$ and $\Uc^\Gamma$, respectively, and the
complicated dependence of $\Uc^M$ and $\Uc^\Gamma$ on $\mu$ gives rise
to an intricate diagram.  Remarkably, it is possible to render the
system nontrivial by tuning \U\ to values of the order of the hopping
$t$.

\emph{Conclusion:} We described symmetries which guarantee the
existence of zero-energy impurity bound states in TR-invariant SCs for
a critical value of the impurity strength.  The same symmetries allow
the definition of the position-space topological \Ztwo\ invariant
\QDIII\ which we related to the bulk \Ztwo\ invariant \QZZ.  The
relevance of our findings was demonstrated for the TR-invariant
$p$-wave phase of the doped KH model, where symmetries protect the
zero-energy crossings and a lattice of impurities can change the bulk
topological order of the system.  Finally, we have shown that TR
invariant topologically non-trivial SCs can be made robust against
low-energy impurity states by strongly breaking all additional
symmetries.  This improves prospects for protocols utilizing
topologically protected Majorana zero-energy states.

\emph{Acknowledgments:} We acknowledge valuable discussions with
E.~Demler, J.~Moore, and B.~Zocher, and would like to thank
G.~Khaliullin for collaboration in an early stage of this work.  L.K.
acknowledges financial support by ESF and T.H. by the Dutch Science
Foundation NWO/FOM. This work was supported in part by the National
Science Foundation under Grant No. PHYS-1066293 and the hospitality of
the Aspen Center for Physics.

\clearpage
\onecolumngrid

\section*{Supplemental Material}
\label{sec:SupplementalMaterial}

\setcounter{equation}{7}
\setcounter{table}{1}

\subsection{Accidental zero-energy crossings in class DIII}
In this section we show that besides the symmetries which make the
phase of $\pf(D)$ independent of the impurity strenght,  there are other sufficient
conditions which yield the same result, although they are not related
to symmetries.  We believe that these conditions are less relevant,
because they will probably not continue to hold when the order
parameter is calculated self-consistently.  Moreover, we utilize again
the TR invariant $p$-wave phase of the doped KH model to demonstrate
that the symmetry unrelated conditions do matter when one investigates
theoretically whether certain mean field Hamiltonians can have
impurity induced zero-energy states.

Observe first that
\begin{equation}
  \pf \begin{pmatrix}
    \alpha A & C\\
    -C^T & \beta B
  \end{pmatrix} = \pf \begin{pmatrix}
    \beta A & C\\
    -C^T & \alpha B
  \end{pmatrix} \label{eq:PfaffianProperty}
\end{equation}
for arbitrary $\alpha,\beta\in\mathbb{C}$ and $A,B,C$ complex $N\times N$
matrices with $A=-A^T,B=-B^T$, which follows directly from the definition of
the Pfaffian
\begin{equation}
  \pf(D) \equiv \frac{1}{2^NN!} \sum_{\sigma\in S_{2N}} \sgn(\sigma)
  \underbrace{ \prod_{i=1}^{N} D_{\sigma(2i-1),\sigma(2i)} }_{P_{D,\sigma}}
\end{equation}
where  $P_{D,\sigma}$ defined above is only a function of the
product $\alpha\beta$ but not of $\alpha$ or $\beta$ indepenently.  Moreover,
in the presence of TR symmetry one has $h=\sigma_2 h^T \sigma_2$ and
$\Delta=\sigma_2\Delta^*\sigma_2$, so that one can write
\begin{equation}
  \begin{split}
    [\pf(D)]^* &= \pf[(h\sigma_2+\Delta)^*] \\
    &= \det(i\sigma_2)\pf(-h^T\sigma_2 + \Delta^*)\\
    &= \pf(-h \sigma_2 + \Delta), \label{eq:PfDccMainText}
  \end{split}
\end{equation}
where we used $\det(i\sigma_2)=1$ and the identity $\det(B)\pf(A) =
\pf(BAB^T)$.  Starting from this equation, one can make use of
Eq.~\eqref{eq:PfaffianProperty} to show that any of the four
conditions
\begin{enumerate}
\item $ih_1+h_2=c(-d_1+id_2)$, $d_0=d_3=\bm{0}_N$,
\item $ih_1+h_2=c(-d_1+id_2)$, $h_0=h_3=\bm{0}_N$,
\item $d_1=cd_2$, $h_1=h_2=h_3=d_0=\bm{0}_N$,
\item $h_1=ch_2$, $h_3=d_0=d_1=d_2=\bm{0}_N$,
\end{enumerate}
suffices to have $[\pf(D)]^*=\pm\pf(D)$ where $c\in\mathbb{R}$.  If,
for example, the first condition is satisfied one may write
\begin{align}
\begin{split}
  [\pf(D)]^* &= (-1)^N\det\sigma_3 \pf \begin{pmatrix}
    (-d_1+id_2)(1-c) & ih_0 + ih_3 \\
    -ih_0 + ih_3 & (d_1 + id_2)(1+c)
  \end{pmatrix}\\
  &= (-1)^N\pf \begin{pmatrix}
    (-d_1+id_2)(1-c) & -ih_0 - ih_3 \\
    ih_0 - ih_3 & (d_1 + id_2)(1+c)
  \end{pmatrix}\\
  &= (-1)^N \pf \begin{pmatrix}
    (-d_1+id_2)(1+c) & -ih_0 - ih_3 \\
    ih_0 - ih_3 & (d_1 + id_2)(1-c)
  \end{pmatrix}\\
  &= (-1)^N\pf(D).
\end{split}
\end{align}
We revisit the example from the main text where  a Rashba spin-orbit coupling term was added to the mean field
Hamiltonian.  Since for
finite $\lambda_\mathrm{R}$ the zero-energy states are protected only
by spatial symmetries one expects any spatially random perturbation to
cause avoidance of the impurity induced zero-energy crossing.  Hence,
it is surprising to find that for $\lambda_\mathrm{R}>0$,
$\kappa_\gamma=1$, the presence of spatially random entries in the matrices $h_0$ and $h_3$ does
not lead to  avoided  zero-energy crossings.  This finding can  be
understood by observing that the $\vec{d}$ vector and the isotropic
Rashba coupling in position space obey the relation
$ih_1+h_2=c(-d_1+id_2)$ while $d_0=d_3=0$ which is the first of the
four conditions stated above.  However, when choosing 
$\kappa_x = \kappa_y \neq \kappa_z$,  this condition is violated,  and in this case only
such nonmagnetic disorder which is compatible with the remaining $M_z$
mirror symmetry preserves the zero-energy crossing.

\subsection{Coupling of symmetry-related states}
In this section we show explicitly that states which are related
through one of the symmetry operators $P=\tau_1 K,\ T=i\sigma_2 K,\
C=\tau_1\otimes\sigma_2$ can ($C$) or cannot ($T,P$) be coupled to
each other by a TR invariant BdG Hamiltonian.  The state vector under
consideration is written as $\ket{\psi} = \left( \begin{smallmatrix} u
    \\ v
  \end{smallmatrix}\right)$.

\paragraph{PH symmetry}
\begin{equation}
  \begin{split}
    \braket{\psi}{{\cal H}P}{\psi}
    &= \begin{pmatrix}u^\dagger&v^\dagger\end{pmatrix}
    \begin{pmatrix} h & \Delta \\
      -\Delta^* & -h^T
    \end{pmatrix}\begin{pmatrix}
      v^* \\ u^*
    \end{pmatrix} = u^\dagger h v^* - v^\dagger h^T u^* +
    u^\dagger\Delta u^* - v^\dagger\Delta^* v^* = 0
  \end{split}
\end{equation}
where the last equality holds because $u^\dagger h v^* = v^\dagger
h^Tu^*$ and $\Delta^T=-\Delta$.

\paragraph{TR symmetry}
\begin{equation}
  \begin{split}
    \braket{\psi}{{\cal H}T}{\psi}
    &= \begin{pmatrix}u^\dagger &v^\dagger \end{pmatrix} 
    \begin{pmatrix} h & \Delta \\
      -\Delta^* & -h^T
    \end{pmatrix}\begin{pmatrix}
      i\sigma_2 u^* \\ i\sigma_2 v^*
    \end{pmatrix}
    = i\left( u^\dagger h\sigma_2 u^* + u^\dagger \Delta\sigma_2 v^*  -
      v^\dagger \Delta^*\sigma_2 u^* - v^\dagger h^T\sigma_2 v^* \right)
    = 0
  \end{split}
\end{equation}
where we used the TR invariance of the Hamiltonian which implies that
$h\sigma_2$ and $h^T\sigma_2$ are antisymmetric matrices and that
$u^\dagger \Delta\sigma_2 v^* = v^\dagger \Delta^*\sigma_2 u^*$.

\paragraph{Chiral symmetry}
\begin{equation}
  \begin{split}
    \braket{\psi}{{\cal H}C}{\psi}
    &= \begin{pmatrix}u^\dagger &v^\dagger \end{pmatrix}
    \begin{pmatrix} h & \Delta \\
      -\Delta^* & -h^T
    \end{pmatrix}\begin{pmatrix}
      \sigma_2 v \\ \sigma_2 u
    \end{pmatrix}
    = u^\dagger h\sigma_2 v + u^\dagger \Delta\sigma_2 u  -
      v^\dagger \Delta^*\sigma_2 v - v^\dagger h^T\sigma_2 u
    \neq 0
  \end{split}
\end{equation}
This demonstrates that $\ket{\psi}$ and $C\ket{\psi}$ in general can
be coupled by TR invariant BdG Hamiltonians.  In the main text we
showed, that this coupling must vanish if there is a symmetry $U$ of
the Hamiltonian that anticommutes with $C$.

\subsection{Explanation of Eq.~(4)}
In this section we explain why every unitary operator $U$ which
anticommutes with the chiral symmetry operator and is compatible with
the PH redundancy in Eq.~(1) automatically obeys Eq.~(4).

Because of the PH redundancy in the Hamiltonian Eq.~(1), the
general unitary transformation $f_\alpha =
\sum_{\beta=1}^{2N}(s_{\alpha\beta}c_\beta +
t_{\alpha\beta}c_\beta^\dagger)$ reads in matrix form 
\begin{equation}
  U = \begin{pmatrix}
    s & t \\ t^* & s^*
  \end{pmatrix} 
  \Leftrightarrow
  [U,P]=0,
\end{equation}
i.e. every $4N \times 4N$ transformation matrix $U$, which respects
the PH redundancy of the Hamiltonian and thus is physically
meaningful, has to commute with the PH operator $P=\tau_1 K$.  On the
other hand, the most general unitary operator anticommuting with the
chiral symmetry operator $C$ reads
\begin{equation}
  U =  \frac{1}{2} \begin{pmatrix}
    u_1\sigma_2 + \sigma_2u_2 & u_1 - \sigma_2
    u_2 \sigma_2 \\
    u_2 - \sigma_2 u_1 \sigma_2 & -\sigma_2 u_1 - u_2\sigma_2
  \end{pmatrix}
  \Leftrightarrow
  V^\dagger U V = \begin{pmatrix}
    \bm{0}_{2N} & u_1 \\
    u_2 & \bm{0}_{2N}
  \end{pmatrix}.
\end{equation}
Hence one infers that every unitary operator $U$ which anticommutes
with $C$ and is compatible with the PH redundancy in Eq.~(1) is of the
form
\begin{equation}
  U =  \frac{1}{2} \begin{pmatrix}
    W\sigma_2 + \sigma_2W^* & W - \sigma_2
    W^* \sigma_2 \\
    W^* - \sigma_2 W \sigma_2 & -\sigma_2 W - W^*\sigma_2
  \end{pmatrix}
  \Leftrightarrow
  V^\dagger U V = \begin{pmatrix}
    \bm{0}_{2N} & W \\
    W^* & \bm{0}_{2N}
  \end{pmatrix}.
\end{equation}


\begin{thebibliography}{99}
\bibitem{Balatsky2006} A. V. Balatsky, I. Vekhter, and Jian-Xin Zhu,
  \href{http://link.aps.org/doi/10.1103/RevModPhys.78.373}{Rev. Mod. Phys. {\bf
      78}, 373 (2006)}.
\bibitem{Alloul2009} H. Alloul, J. Bobroff, M. Gabay, and
  P. J. Hirschfeld,
  \href{http://link.aps.org/doi/10.1103/RevModPhys.81.45}{Rev. Mod. Phys. 81,
    45 (2009)}.
\bibitem{Hudson2001} E. W. Hudson, K. M. Lang, V. Madhavan, S. H. Pan,
  H. Eisaki, S. Uchida, and J. C. Davis,
  \href{http://dx.doi.org/10.1038/35082019}{Nature (London) {\bf 411},
    920 (2001)}.
\bibitem{Sau2012} J. D. Sau, and E. Demler,
  \href{http://dx.doi.org/10.1103/PhysRevB.88.205402}{Phys. Rev. B
    {\bf 88}, 205402 (2013)}.
\bibitem{Hu2013} H. Hu, L. Jiang, H. Pu, Y. Chen, and X.-J. Liu,
  \href{http://link.aps.org/doi/10.1103/PhysRevLett.110.020401}{
    Phys. Rev. Lett. {\bf 110}, 020401 (2013)}.
\bibitem{Maki1999} K. Maki and E. Puchkaryov,
  \href{http://dx.doi.org/10.1209/epl/i1999-00156-5}{Europhys. Lett. {\bf
      45}, 263 (1999)}.
\bibitem{Khaliullin1997} G. Khaliullin, R. Kilian, S. Krivenko, and
  P. Fulde,
  \href{http://link.aps.org/doi/10.1103/PhysRevB.56.11882}{Phys. Rev. B
    {\bf 56}, 11882 (1997)}.
\bibitem{Abrikosov1960} A. A. Abrikosov and L. P. Gor'kov,
  Zh. Eksperim. i Teor. Fiz. {\bf 39}, 1781 (1960) [Soviet Phys.-JETP
  12, 1243 (1961)].
\bibitem{Anderson1959} P. W. Anderson,
  \href{http://dx.doi.org/10.1016/0022-3697(59)90036-8}{J. Phys. Chem. Solids
    {\bf 11}, 26 (1959)}.
\bibitem{Altland1997} A. Altland and M. R. Zirnbauer,
  \href{http://link.aps.org/doi/10.1103/PhysRevB.55.1142}{Phys. Rev. B {\bf
      55}, 1142 (1997)}; M. R. Zirnbauer,
  \href{http://dx.doi.org/10.1063/1.531675}{J. Math. Phys. {\bf 37}, 4986
    (1996)}.
\bibitem{Schnyder2008} A. P. Schnyder, S. Ryu, A. Furusaki, and
  A. W. W. Ludwig,
  \href{http://link.aps.org/doi/10.1103/PhysRevB.78.195125}{Phys. Rev. B {\bf
      78}, 195125 (2008)}.
\bibitem{Kitaev2009} A. Kitaev,
  \href{http://arxiv.org/abs/0901.2686}{arXiv:0901.2686 [cond-mat.mes-hall]}.
\bibitem{Qi2009} X.-L. Qi, T. L. Hughes, S. Raghu, and S.-C. Zhang,
  \href{http://link.aps.org/doi/10.1103/PhysRevLett.102.187001}{Phys. Rev. Lett. 102,
    187001 (2009)}.
\bibitem{Hasan2010} M. Z. Hasan and C. L. Kane,
  \href{http://link.aps.org/doi/10.1103/RevModPhys.82.3045}{Rev. Mod. Phys. {\bf
      82}, 3045 (2010)}.
\bibitem{Qi2011} X.-L. Qi and S.-C. Zhang,
  \href{http://link.aps.org/doi/10.1103/RevModPhys.83.1057}{Rev. Mod. Phys. {\bf
      83}, 1057 (2011)}.
\bibitem{Sato2009} M. Sato,
  \href{http://link.aps.org/doi/10.1103/PhysRevB.79.214526}{Phys. Rev. B {\bf
      79}, 214526 (2009)}.
\bibitem{Fu2010} L. Fu and E. Berg,
  \href{http://link.aps.org/doi/10.1103/PhysRevLett.105.097001}{Phys. Rev. Lett. {\bf
      105}, 097001 (2010)}.
\bibitem{footnote:invariant3D} {More generally, the topological
    invariant in three dimensions is a $\mathbb{Z}$-invariant.
    \cite{Schnyder2008} \QZZ\ is the parity of this
    $\mathbb{Z}$-invariant.\cite{Sato2009, Fu2010}}
\bibitem{Read2000} N. Read and D. Green,
  \href{http://link.aps.org/doi/10.1103/PhysRevB.61.10267}{Phys. Rev. B {\bf
      61}, 10267 (2000)}.
\bibitem{Kitaev2001} A. Y. Kitaev,
  \href{http://dx.doi.org/10.1070/1063-7869/44/10S/S29}{Physics-Uspekhi
    {\bf 44}, 131 (2001)}.
\bibitem{Ivanov2001} D. A. Ivanov,
  \href{http://link.aps.org/doi/10.1103/PhysRevLett.86.268}{Phys. Rev. Lett. {\bf
      86}, 268 (2001)}.
\bibitem{Wilczek2009} F. Wilczek,
  \href{http://dx.doi.org/10.1038/nphys1380}{Nature Physics 5, 614 (2009)}.
\bibitem{Fulga2013} I. C. Fulga, B. van Heck, M. Burrello, T. Hyart,
  \href{http://dx.doi.org/10.1103/PhysRevB.88.155435}{Phys. Rev. B
    {\bf 88}, 155435 (2013)}.
\bibitem{Kopnin1991} N. B. Kopnin and M. M. Salomaa,
  \href{http://link.aps.org/doi/10.1103/PhysRevB.44.9667}{Phys. Rev. B
    {\bf 44}, 9667 (1991)}.
\bibitem{Volovik1999} G. E. Volovik,
  \href{http://dx.doi.org/10.1134/1.568223}{JETP Letters {\bf 70}, 609
    (1999)}.
\bibitem{footnote:classAII} {The normal state single-particle
    Hamiltonian $H_0(u)$ in the limit $u\rightarrow -\infty$ has one
    additional Kramers pair occupied compared to the limit
    $u\rightarrow +\infty$.  Thus, upon continuously tuning $u$ from
    large negative to large positive values, at least one Kramers pair
    has to cross the Fermi level at zero energy.}
\bibitem{SupplementalMaterial} See Supplemental Material for details
  about matrix elements of $H_\Delta$, Eq. (4), and zero-energy
  crossings not protected by symmetry.
\bibitem{Kitaev2006} A. Kitaev,
  \href{http://dx.doi.org/10.1016/j.aop.2005.10.005}{Ann. Phys. {\bf
      321}, 2 (2006)}.
\bibitem{Giniyat0910} G. Jackeli and G. Khaliullin,
  \href{http://link.aps.org/doi/10.1103/PhysRevLett.102.017205}{Phys. Rev. Lett. {\bf
      102}, 017205 (2009)}; J. Chaloupka, G. Jackeli and
  G. Khaliullin,
  \href{http://link.aps.org/doi/10.1103/PhysRevLett.105.027204}{Phys. Rev. Lett. {\bf
      105}, 027204 (2010)}.
\bibitem{Ueno2013} Y. Ueno, A. Yamakage, Y. Tanaka, and M. Sato,
  \href{http://dx.doi.org/10.1103/PhysRevLett.111.087002}{Phys. Rev. Lett. {\bf
      111}, 087002 (2013)}.
\bibitem{Zhang2013} F. Zhang, C. L. Kane, and E. J. Mele,
  \href{http://dx.doi.org/10.1103/PhysRevLett.111.056403}{Phys. Rev. Lett. {\bf
    111}, 056403 (2013)}.
\bibitem{Chiu2013} C.-K. Chiu, H. Yao, and S. Ryu,
  \href{http://dx.doi.org/10.1103/PhysRevB.88.075142}{Phys. Rev. B
    {\bf 88}, 075142 (2013)}.
\bibitem{Marimoto2013} T. Morimoto and A. Furusaki,
  \href{http://dx.doi.org/10.1103/PhysRevB.88.125129}{Phys. Rev. B
    {\bf 88}, 125129 (2013)}.
\bibitem{footnote:gaplessTopSC} {The gap can close at momenta away from the
    TRIM, such that \QDIII\ cannot be defined whereas \QZZ\ still is
    well-defined, cf. \eqref{eq:DefinitionQZZ}.  This indicates gapless
    topological superconductivity.\cite{Sato2010}}
\bibitem{Sato2010} M. Sato and S. Fujimoto,
  \href{http://link.aps.org/doi/10.1103/PhysRevLett.105.217001}{Phys. Rev. Lett. {\bf
      105}, 217001 (2010)}.
\bibitem{Hyart2012} T. Hyart, A. R. Wright, G. Khaliullin, and
  B. Rosenow,
  \href{http://link.aps.org/doi/10.1103/PhysRevB.85.140510}{Phys. Rev. B
    {\bf 85}, 140510 (2012)}.
\bibitem{Okamoto2012} S. Okamoto,
  \href{http://link.aps.org/doi/10.1103/PhysRevB.87.064508}{Phys. Rev. B {\bf
      87}, 064508 (2013)}.
\bibitem{Scherer2014} D. D. Scherer, M. M. Scherer, G. Khaliullin,
  C. Honerkamp, and B. Rosenow,
  \href{http://dx.doi.org/10.1103/PhysRevB.90.045135}{Phys. Rev. B {\bf 90},
    045135 (2014)}.
\bibitem{Hyart2014} T.~Hyart, A.~R.~Wright, and B.~Rosenow,
  \href{http://dx.doi.org/10.1103/PhysRevB.90.064507}{Phys. Rev. B
    {\bf 90}, 064507 (2014)}.
\bibitem{You2012} Y.-Z. You, I. Kimchi, and A. Vishwanath,
  \href{http://link.aps.org/doi/10.1103/PhysRevB.86.085145}{Phys. Rev. B {\bf
      86}, 085145 (2012)}.

\end{thebibliography}
\end{document}